\documentstyle[amsfonts,aps,twocolumn]{revtex}
\input epsf

\begin{document}
\draft
\title{\begin{flushright}
\footnotesize{CECS-PHY-00/01}\\
\footnotesize{ULB-TH-00/01}
\end{flushright}{\vskip 1.5cm} Black Hole Scan {\vskip 1cm}}

\author{Juan Cris\'{o}stomo$^{1,2}$, Ricardo Troncoso$^{1,4}$ and Jorge Zanelli$%
^{1,3}$}
\address{$^{1}$Centro de Estudios Cient\'{\i }ficos, CECS, Casilla 1469, Valdivia,
Chile.\\ $^{2}$Departamento de F\'{\i }sica, Facultad de Ciencias,
Universidad de Chile, Casilla 653, Santiago, Chile. \\
$^{3}$Departamento de F\'{\i }sica, Universidad de Santiago de
Chile, Casilla 307, Santiago 2, Chile.\\ $^{4}$Physique
Th\'{e}orique et Math\'{e}matique, Universit\'{e} Libre de
Bruxelles, Campus Plaine, C.P.231, B-1050, Bruxelles, Belgium.}
\maketitle

\begin{abstract}
\hspace*{3cm} {\bf Abstract} \vskip 0.6cm  Gravitation theories
selected by requiring that they have a unique anti-de Sitter
vacuum with a fixed cosmological constant are studied. For a
given dimension $d$,\ the Lagrangians under consideration are
labeled by an integer $k=1,2,...,\left[ \frac{d-1}{2}\right] $.
Black holes for each $d$ and $k$ are found and are used to rank
these theories. A minimum possible size for a localized
electrically charged source is predicted in the whole set of
theories, except General Relativity.

It is found that the thermodynamic behavior falls into two classes: If $%
d-2k=1$, these solutions resemble the three dimensional black hole;
otherwise, their behavior is similar to the Schwarzschild-AdS$_{4}$ geometry.

\end{abstract}

\section{Introduction}

Black holes are much more than a particular class of exact solutions of the
Einstein Equations; they are an essential feature of the spacetime dynamics
in almost any sensible theory of gravity. Within the framework of General
Relativity, the singularity theorems of Hawking and Penrose \cite
{Hawking-Penrose} show that singular configurations --such as the
Schwarzschild black hole-- are inevitable under quite generic initial
conditions. Furthermore, the Schwarzschild solution describes the leading
asymptotic behavior of the geometry for any localized distribution of
matter.\ The existence of this solution at spacelike infinity is a central
ingredient to prove the positivity of energy in General Relativity \cite
{Witten-Energy}. On the other hand, black holes are also fundamental objects
where the thermodynamics of the gravitational field and its connection with
information theory is expected to shed light on the quantization problem.

In this paper, we survey the black hole solutions in a class of gravitation
theories, selected by requiring that they have a unique anti-de Sitter
vacuum with a fixed cosmological constant. For a given dimension $d$,\ the
Lagrangians under consideration are labeled by an integer $k=1,2,...,\left[
\frac{d-1}{2}\right] $, where the Einstein-Hilbert Lagrangian corresponds to
$k=1$.\ For each of these theories we examine their static, spherically
symmetric solutions. The existence of physical black holes is then used as a
criterion to assess the validity of those theories, leading to a natural
splitting between theories with even and odd $k$.

Coupling these gravity theories with the Maxwell action predicts the
smallest size of a spherically symmetric electrically charged source, except
for $k=1$.

An important aspect of the black holes under consideration is their
thermodynamics, which is expected to be a reflection of the underlying
quantum theory. The canonical ensemble for minisuperspaces containing the
black holes found in these theories is well defined provided a negative
cosmological constant exists. It is found that black holes are unstable
against decay by Hawking radiation, unless their horizon radius is large,
compared to the AdS radius.

Among all theories under consideration, there is only one representative in
each odd dimension, given by a Chern-Simons action, having physical black
holes whose spectrum has a mass gap separating them from AdS spacetime.
These black holes always reach thermal equilibrium with a heat bath, and
have positive specific heat, which guarantees their stability under thermal
fluctuations.

\subsection{Higher Dimensional Gravity Revisited}

The standard higher dimensional extension of the four-dimensional
Einstein-Hilbert ({\bf EH}) action reads \cite{GTilde}

\begin{equation}
I_{EH}=-\frac{1}{2(d-2)\Omega _{d-2}G}\int d^{d}x\sqrt{-g}(R-2\Lambda ).
\label{EH}
\end{equation}
String and $M$-theory corrections to this action would bring in higher
powers of curvature --see, e.g. Refs. \cite{CHSW,GV}. This may be a source
of inconsistencies because higher powers of curvature could give rise to
fourth order differential equations for the metric. This not only
complicates the causal evolution, but in general would introduce ghosts and
violate unitarity. However, Zwiebach \cite{Zwiebach} and Zumino \cite{Zumino}
observed that ghosts are avoided if stringy corrections would only consist
of the dimensional continuations of the Euler densities, so that the
resulting field equations remain second order.

These theories are far from exotic. Indeed, they are described by the most
general Lagrangians constructed with the same principles as General
Relativity, that is, general covariance and second order field equations for
the metric. These theories were first discussed by Lanczos for $d=5$ in $%
1938 $ \cite{Lanczos} and more recently by Lovelock for $d\geq 3$ \cite
{Lovelock}.

The Lanczos-Lovelock ({\bf LL}) action is a polynomial of degree $[d/2]$ in
curvature\footnote{%
Here $[x]$ is the integer part of $x$.}, which can also be written in terms
of the Riemann curvature $R^{ab}=d\omega ^{ab}+\omega _{c}^{a}\omega ^{cb}$
and the vielbein $e^{a}$ as\footnote{%
Wedge product between forms is understood throughout.}

\begin{equation}
I_{G}=\kappa \int \sum_{p=0}^{[d/2]}\alpha _{p}L^{(p)},  \label{Lovaction}
\end{equation}
where $\alpha _{p}$ are arbitrary constants, and $L^{(p)}$ is given by

\begin{equation}
L^{(p)}=\epsilon _{a_{1}\cdots a_{d}}R^{a_{1}a_{2}}\!\cdot \!\cdot \!\cdot
\!R^{a_{2p-1}a_{2p}}e^{a_{2p+1}}\!\cdot \!\cdot \!\cdot \!e^{a_{d}}.
\label{Lovlag}
\end{equation}
In first order formalism the action (\ref{Lovaction}) is regarded as a
functional of the vielbein and the spin connection, and the corresponding
field equations obtained varying with respect to $e^{a}$ and $\omega ^{ab}$%
read
\begin{eqnarray}
\sum_{p=0}^{[\frac{d-1}{2}]}\alpha _{p}(d-2p){\cal E}_{a}^{p} &=&0\;,
\label{E-L} \\
\sum_{p=1}^{[\frac{d-1}{2}]}\alpha _{p}p(d-2p){\cal E}_{ab}^{p} &=&0\;,
\label{Tor}
\end{eqnarray}
where we have defined

\begin{eqnarray*}
{\cal E}_{a}^{p}:= &&\epsilon _{ab_{1}\cdots b_{d-1}}R^{b_{1}b_{2}}\cdots
R^{b_{2p-1}b_{2p}}e^{b_{2p+1}}\cdots e^{b_{d-1}},\hspace{-0.06in} \\
{\cal E}_{ab}^{p}:= &&\epsilon _{aba_{3}\cdots a_{d}}R^{a_{3}a_{4}}\cdots
R^{a_{2p-1}a_{2p}}T^{a_{2p+1}}e^{a_{2p+2}}\cdots e^{a_{d}}.\hspace{-0.06in}
\end{eqnarray*}
Here $T^{a}=de^{a}+\omega _{b}^{a}e^{b}$ is the torsion $2$-form.

Note that in even dimensions, the term $L^{(d/2)}$\ is the Euler density and
therefore does not contribute to the field equations. However, the presence
of this term in the action --with a fixed weight factor-- guarantees the
existence of a well defined variational principle for asymptotically locally
AdS spacetimes \cite{ACOTZ3+1,ACOTZ2n}. Moreover, the Euler density should
assign different weights to non-homeomorphic geometries in the quantum
theory.

The first two terms in the LL action (\ref{Lovaction})\ are the cosmological
and kinetic terms of the EH action (\ref{EH}) respectively, and therefore
General Relativity is contained in the LL theory as a particular case.

The linearized approximation of the LL and EH actions around a flat,
torsionless background are classically equivalent \cite{Zumino}. However,
beyond perturbation theory the presence of higher powers of curvature in the
Lagrangian makes both theories radically different. In particular, black
holes and big-bang solutions of (\ref{Lovaction}), have different asymptotic
behaviors from their EH counterparts in general. Hence, a generic solution
of the LL action cannot be approximated by a solution of Einstein's theory.

\subsection{Drawbacks}

For a given dimension and an arbitrary choice of coefficients $\alpha _{p}$%
's, higher dimensional LL theories have some drawbacks. One difficulty is
the fact that the dynamical evolution can become unpredictable because the
Hessian matrix cannot be inverted for a generic field configuration. Thus,
the velocities are multivalued functions of the momenta and therefore the
passage from the Lagrangian to the Hamiltonian is ill defined \cite{tz,HTZ}.

A reflection of this problem can be viewed in the static, spherically
symmetric solutions of (\ref{E-L}) and (\ref{Tor}). For arbitrary $\alpha
_{p}$'s there are negative energy solutions with horizons and positive
energy solutions with naked singularities \cite{Wheeler}.

These problems can be curbed if the coefficients $\alpha _{p}$'s are chosen
in a suitable way. The aim of the next section is to show that requiring the
theories to possess {\em a unique }cosmological constant, strongly restricts
the coefficients $\alpha _{p}$'s. As a consequence, one obtains a set of
theories labelled by an integer $k$ which lead to well defined black hole
configurations.

\section{Selecting Sensible Theories}

The field equations of LL theory (\ref{E-L}) can be rearranged as a
polynomial of $k$th degree in the curvature

\begin{equation}
\epsilon _{ab_{1}\cdots b_{d-1}}\beta _{0}\bar{R}_{\beta
_{1}}^{b_{1}b_{2}}\cdot \cdot \cdot \bar{R}_{\beta
_{k}}^{b_{2k-1}b_{2k}}e^{b_{2k+1}}\!\cdot \cdot \cdot e^{b_{d-1}}=0
\label{LoveBeta}
\end{equation}
where $\bar{R}_{\beta _{i}}^{ab}:=R^{ab}+\beta _{i}e^{a}e^{b}$, and the
coefficients $\beta _{i}$'s are related to the $\alpha _{p}$'s through
\begin{equation}
\sum\limits_{p}^{[\frac{d-1}{2}]}(d-2p)\alpha _{p}x^{p}=\beta
_{0}\prod\limits_{i}^{k}(x-\beta _{i}).  \label{Poly}
\end{equation}

Equation (\ref{LoveBeta}) can possess in general, several constant curvature
solutions with different radii\ $r_{i}=|\beta _{i}|^{-1/2}$, making the
value of the cosmological constant ambiguous. In fact, the cosmological
constant could change in different regions of a spatial section, or it could
jump arbitrarily as the system evolves in time \cite{tz,HTZ}.

On the other hand, solving (\ref{LoveBeta}) for a given global isometry
leads in general to several solutions with different asymptotic behaviors.
Some of these solutions are ``spurious'' in the sense that perturbations
around them yield ghosts. For instance, if $\alpha _{1}$ and $\alpha _{2}$
were the only nonvanishing coefficients in the LL action (\ref{Lovlag}), two
different static, spherically symmetric solutions would be obtained, which
are asymptotically (A)dS and flat respectively.\ The perturbations around
the latter solution are gravitons, while those on the former are spurious in
the sense described above \cite{Boulware-Deser}.

These problems are overcome demanding the theory to have {\em a unique}
cosmological constant.

Requiring the existence of {\em a unique} cosmological constant implies that
locally maximally symmetric solutions possess only one fixed radius, that is
$R^{ab}=-\beta e^{a}e^{b}$. This in turn means that the polynomial (\ref
{Poly}) must have only one real root. Hence, the coefficients $\alpha _{p}$%
's are fixed through equation (\ref{Poly}), so that the real $\beta $'s in (%
\ref{LoveBeta}) are all equal, allowing --for $d\geq 7$-- an arbitrary
number of distinct imaginary $\beta $'s which must come in conjugate pairs.
Under this assumption, solutions representing localized sources of matter
approach a constant curvature spacetime with a fixed radius in the
asymptotic region.

In what follows, we consider the simplest class of such theories, namely, we
assume the field equations to be of the form (\ref{LoveBeta}) with only one
real $\beta :=\frac{1}{l^{2}}$, and no complex roots\footnote{%
A negative cosmological constant is assumed for later convenience, but this
analysis does not depend on its sign.}. These theories are described by the
action
\begin{equation}
I_{k}=\kappa \int \sum_{p=0}^{k}c_{p}^{k}L^{(p)}\;,  \label{Ik}
\end{equation}
which is obtained from (\ref{Lovaction}) with the choice

\begin{equation}
\alpha _{p}:=c_{p}^{k}=\left\{
\begin{array}{ll}
\frac{l^{2(p-k)}}{(d-2p)}\left(
\begin{array}{c}
k \\
p
\end{array}
\right)  & ,\text{ }p\leq k \\
0 & ,\text{ }p>k
\end{array}
\right.   \label{Coefs}
\end{equation}
where $1\leq k\leq [\frac{d-1}{2}]$.

For a given dimension $d$, the coefficients $c_{p}^{k}$ give rise to a
family of inequivalent theories, labeled by the integer $k\in \{1,...,[\frac{%
d-1}{2}]\}$ which represents the highest power of curvature in the
Lagrangian. This set of theories possesses only two fundamental constants, $%
\kappa $ and $l$, related to the gravitational constant $G_{k}$ and the
cosmological constant $\Lambda $ through\footnote{%
Here the gravitational constant has natural units given by $[G_{k}]=($length$%
)^{d-2k}$.}

\begin{eqnarray}
\kappa &=&\frac{1}{2(d-2)!\Omega _{d-2}G_{k}},  \label{Kappa} \\
\Lambda &=&-\frac{(d-1)(d-2)}{2l^{2}}.  \label{Lambda}
\end{eqnarray}

The field equations for the action $I_{k}$ in (\ref{Ik}), read
\begin{eqnarray}
\epsilon _{ba_{1}\cdots a_{d-1}}\bar{R}^{a_{1}a_{2}}\!\cdot \!\cdot \!\cdot
\!\bar{R}^{a_{2k-1}a_{2k}}e^{a_{2k+1}}\!\cdot \!\cdot \!\cdot \!e^{a_{d-1}}
&=&0,  \label{kEinstein} \\
\epsilon _{aba_{3}\cdots a_{d}}\bar{R}^{a_{3}a_{4}}\!\cdot \!\cdot \!\cdot \!%
\bar{R}^{a_{2k-1}a_{2k}}T^{a_{2k+1}}e^{a_{2k+2}}\!\cdot \!\cdot \!\cdot
\!e^{a_{d-1}} &=&0,  \label{kTorsion}
\end{eqnarray}
with $\bar{R}^{ab}:=R^{ab}+\frac{1}{l^{2}}e^{a}e^{b}$.

\subsection{Examples}

There are special cases of interest which are obtained for particular values
of the integer $k$.

\begin{itemize}
\item  The Einstein-Hilbert action in $d$ dimensions (\ref{EH}) is recovered
setting $k=1$ in (\ref{Ik}).

\item  At the other end of the range, $k=[\frac{d-1}{2}]$, even and odd
dimensions must be distinguished. These cases are exceptional in that they
are the only ones which allow sectors with non-trivial torsion \cite{HDG},
as discussed in Appendix A. When $d=2n-1$, the maximum value of $k$ is $n-1$%
, and the corresponding Lagrangian is a Chern-Simons ({\bf CS}) $2n-1$-form
defined through (\ref{dCS}). For $d=2n$ and $k=n-1$, the action can be
written as the Pfaffian of the $2$-form $\bar{R}^{ab}=R^{ab}+\frac{1}{l^{2}}%
e^{a}e^{b}$ and, in this sense, it has a Born-Infeld-like ({\bf BI}-like)
form given by (\ref{BI})\footnote{%
Strictly speaking one must add the Euler density to the Lagrangian in (\ref
{Ik}) with the coefficient $\alpha _{n}=c_{n}^{n-1}:=\frac{l^{2}}{2n}$,
which does not modify the field equations. Therefore, the same BI Lagrangian
(\ref{BI}) is recovered from (\ref{Coefs}) but now the index $p$ ranges from
$0$ to $n$.}.

\item  In three and four dimensions equation (\ref{Coefs}) defines only one
possible theory which corresponds to EH. As is well known, the EH action is
equivalent to CS theory in three dimensions \cite{Achucarro-Townsend-Witten}%
, and for $d=4$\ the EH action coincides with the BI action up to the Euler
density.

\item  In five and six dimensions, there are only two inequivalent theories
which correspond to $k=1,2$. In five dimensions, $k=1$ represents EH and $k=2
$ leads to CS. For $d=6$,\ one obtain EH and BI respectively.

\item  For $d\geq 7$\ there exist other interesting possibilities which are
neither EH, BI nor CS. For instance, consider the theory given by the action
$I_{k}$ in (\ref{Ik}) with $k=2$, which exists only for dimensions greater
than 4. In this case the Lagrangian reads
\begin{equation}
L=\kappa \left( \frac{l^{-4}}{d}L^{(0)}+\frac{2l^{-2}}{d-2}L^{(1)}+\frac{1}{%
d-4}L^{(2)}\right) ,  \label{GB+form}
\end{equation}
with
\begin{eqnarray}
L^{(0)} &=&\epsilon _{a_{1}\cdots a_{d}}e^{a_{1}}\!\cdot \!\cdot \!\cdot
\!e^{a_{d}}\;,  \label{L0} \\
L^{(1)} &=&\epsilon _{a_{1}\cdots a_{d}}R^{a_{1}a_{2}}\!e^{a_{3}}\!\cdot
\!\cdot \!\cdot \!e^{a_{d}}\;,  \label{L1} \\
L^{(2)} &=&\epsilon _{a_{1}\cdots
a_{d}}R^{a_{1}a_{2}}\!R^{a_{3}a_{4}}\!e^{a_{5}}\!\cdot \!\cdot \!\cdot
\!e^{a_{d}}\;.  \label{L2}
\end{eqnarray}
\end{itemize}

Here $L^{(0)}$ and $L^{(1)}$ are proportional to the standard cosmological
and kinetic terms for the EH\ action, and $L^{(2)}$ is proportional to the
four dimensional Gauss-Bonnet density \cite{EGB},
\begin{equation}
{\frak R}^{2}:=(R_{\mu \nu \alpha \beta }R^{\mu \nu \alpha \beta }-4R_{\mu
\nu }R^{\mu \nu }+R^{2}),  \label{EGB}
\end{equation}
where $R^{\mu \nu \alpha \beta }$, $R^{\mu \nu }$ and $R$ are the Riemann,
Ricci and scalar curvatures, respectively. The action in standard tensor
components reads
\begin{equation}
I_{2}=\frac{-2(d-3)!\kappa }{l^{2}}\!\!\int\limits_{M}\!\!d^{d}x\sqrt{-g}%
\left[ \frac{l^{2}{\frak R}^{2}}{2(d-3)(d-4)}+R-\Lambda \right] ,
\label{GB+}
\end{equation}
with $\Lambda $\ given by (\ref{Lambda}). In sum, the theory with $k=2$ is
described by a Lagrangian which is a linear combination of Gauss-Bonnet
density, the EH Lagrangian and the volume term with fixed weights.

Each of the theories described by $I_{k}$ for all $k$ possesses a unique
cosmological constant. In fact, as is apparent from equations (\ref
{kEinstein}) and (\ref{kTorsion}), spacetimes satisfying $\bar{R}^{ab}=0$\
are the only locally maximally symmetric solutions. This ensures that
localized matter fields give rise to solutions which are asymptotically AdS
spacetimes.

\section{Static and Spherically Symmetric Solutions}

In this section, we test the theories described by $I_{k}$ analyzing their
static, spherically symmetric solutions including their electrically charged
extensions. It is shown that they possess well behaved black holes,
resembling the Schwarzschild-AdS and Reissner-Nordstrom-AdS solutions. The
subset of theories with $k$ odd differ from their even counterparts, because
in the first case there is a unique black hole solution, whereas in the
latter, an additional solution with a naked singularity exists.

\subsection{Pure Gravity}

Consider static and spherically symmetric solutions of equations (\ref
{kEinstein}) and (\ref{kTorsion}) for a fixed value of the label $k$. In
Schwarzschild-like coordinates, the metric can be written as

\begin{equation}
ds^{2}=-N^{2}(r)f^{2}(r)dt^{2}+\frac{dr^{2}}{f^{2}(r)}+r^{2}d\Omega
_{d-2}^{2}.  \label{gspherical}
\end{equation}
Replacing this ansatz in the field equations (\ref{kEinstein}) and (\ref
{kTorsion}) leads to the following equations for $N$ and $f^{2}$ \cite
{Spheritorsion}

\begin{eqnarray}
\frac{dN}{dr} &=&0,  \nonumber \\
\frac{d}{dr}\left( r^{d-1}\left[ F(r)+\frac{1}{l^{2}}\right] ^{k}\right)
&=&0,  \label{Eqnsgrav}
\end{eqnarray}
where the function $F(r)$ is given by
\begin{equation}
F(r)=\frac{1-f^{2}(r)}{r^{2}}.  \label{F(r)}
\end{equation}
Integrating equations (\ref{Eqnsgrav}) yields

\begin{eqnarray}
N &=&N_{\infty },  \nonumber \\
f^{2}(r) &=&1+\frac{r^{2}}{l^{2}}-\sigma \left( \frac{C_{1}}{r^{d-2k-1}}%
\right) ^{1/k},  \label{f2}
\end{eqnarray}
where the integration constant $N_{\infty }$\ relates coordinate time to the
proper time of an observer at spatial infinity and in what follows is chosen
equal to one. Here $\sigma =(\pm 1)^{(k+1)}$, and the integration constant $%
C_{1}$ is identified as
\[
C_{1}=2G_{k}(M+C_{0})\;,
\]
where $M$\ stands for the mass{\bf ,} as is discussed in detail in section\
III.C.

For even $k$, the ambiguity of sign expressed through $\sigma $ in (\ref{f2}%
) implies that there are two possible solutions provided $C_{1}>0$. The
solution with $\sigma =1$ describes a real black hole with {\em a unique }%
event horizon surrounding the singularity at the origin. The solution with $%
\sigma =-1$ has a naked singularity with positive mass.

If $k$ is odd, there is no ambiguity of sign because $\sigma $ cannot be
different from unity, therefore in that case there exists a unique static,
spherically symmetric solution, which corresponds to a black hole with
positive mass.

The black hole mass for any value of $k$ is a monotonically increasing
function of the horizon radius $r_{+}$, which reads
\begin{equation}
M(r_{+})=\frac{r_{+}^{d-2k{\bf -}1}}{2G_{k}}\left( 1+\frac{r_{+}^{2}}{l^{2}}%
\right) ^{k}-C_{0}.  \label{mass(r)}
\end{equation}
The additive constant $C_{0}$\ is chosen so that the horizon shrinks to a
point for\ $M\rightarrow 0$, hence
\begin{equation}
C_{0}=\frac{1}{2G_{k}}\delta _{d-2k,1}\;,  \label{C0}
\end{equation}
which vanishes in all cases except for CS theory.

Summarizing, for a given dimension $d\geq 3$ the full set of $[\frac{d-1}{2}%
] $ inequivalent theories given by the action $I_{k}$ in (\ref{Ik}), possess
asymptotically AdS black hole solutions whose line elements read

\begin{eqnarray}
ds^{2} &=&-\left( 1+\frac{r^{2}}{l^{2}}-\left( \frac{2G_{k}M+\delta _{d-2k,1}%
}{r^{d-2k-1}}\right) ^{1/k}\right) dt^{2}+  \nonumber \\
&&\frac{dr^{2}}{1+\frac{r^{2}}{l^{2}}-\left( \frac{2G_{k}M+\delta _{d-2k,1}}{%
r^{d-2k-1}}\right) ^{1/k}}+r^{2}d\Omega _{d-2}^{2}\;.  \label{BHGeneral}
\end{eqnarray}

One can see from (\ref{BHGeneral}) that for $k=1$, the three dimensional
black hole \cite{BTZ} and Schwarzschild-AdS solutions of the $d$-dimensional
Einstein-Hilbert action with negative cosmological constant are recovered.
The black hole solutions corresponding to BI and CS theories \cite{btz} are
obtained also from (\ref{BHGeneral}) setting $k=[\frac{d-1}{2}]$.

The whole set of black hole metrics given by (\ref{BHGeneral}) share a
common causal structure when $M>0$, which coincides with the familiar one
described by the Penrose diagram of the four dimensional Schwarzschild-AdS
solution. Nevertheless, the presence of the Kronecker delta within the
metrics (\ref{BHGeneral}) signals the existence of two possible black hole
vacua ($M=0$) with different causal structures. The generic case holds for
the whole set of theories except CS, whose line elements are described by (%
\ref{BHGeneral}) with $d-2k\neq 1$, that is

\begin{eqnarray}
ds^{2} &=&-\left( 1+\frac{r^{2}}{l^{2}}-\left( \frac{2G_{k}M}{r^{d-2k-1}}%
\right) ^{1/k}\right) dt^{2}+  \nonumber \\
&&%
%TCIMACRO{
%\dfrac{dr^{2}}{1+\frac{r^{2}}{l^{2}}-\left( \frac{2G_{k}M}{r^{d-2k-1}}\right) ^{1/k}}}
%BeginExpansion
{\displaystyle {dr^{2} \over 1+\frac{r^{2}}{l^{2}}-\left( \frac{2G_{k}M}{r^{d-2k-1}}\right) ^{1/k}}}%
%EndExpansion
+r^{2}d\Omega _{d-2}^{2}\;.  \label{BHk}
\end{eqnarray}
Analogously with the Schwarzschild-AdS metric, this set possesses a
continuous mass spectrum, whose vacuum state is the AdS spacetime. The other
case is obtained only for $d=2n-1$ dimensions, and it is a peculiarity of CS
theories, whose black hole solutions are recovered from (\ref{BHGeneral})
with $k=n-1$, which read
\begin{eqnarray}
ds^{2} &=&-\left( 1+\frac{r^{2}}{l^{2}}-\left( 2G_{n-1}M+1\right) ^{\frac{1}{%
n-1}}\right) dt^{2}+  \nonumber \\
&&%
%TCIMACRO{
%\dfrac{dr^{2}}{1+\frac{r^{2}}{l^{2}}-\left( 2G_{n-1}M+1\right) ^{\frac{1}{n-1}}}}
%BeginExpansion
{\displaystyle {dr^{2} \over 1+\frac{r^{2}}{l^{2}}-\left( 2G_{n-1}M+1\right) ^{\frac{1}{n-1}}}}%
%EndExpansion
+r^{2}d\Omega _{d-2}^{2}\;.  \label{BHCS}
\end{eqnarray}
In that case, the black hole vacuum ($M=0$) differs from AdS spacetime.
Although this configuration has no constant curvature for $d>3$, it
possesses the same causal structure as the three-dimensional zero mass black
hole. Another common feature with $2+1$ dimensions is the existence of a
mass gap between the zero mass black hole and AdS spacetime, where the later
is obtained for $M=-\frac{1}{2G_{n-1}}$.

\subsection{Coupling to the Electromagnetic Field}

The standard coupling with the electromagnetic field is obtained adding to
the gravitational action $I_{k}$ in Eq. (\ref{Ik}) the Maxwell term\footnote{%
The constant $\epsilon $ is related with the ``vacuum permeability'' through
$\epsilon =\frac{1}{\Omega _{d-2}\epsilon _{0}}$. Its natural units are $%
[\epsilon ]=($length$)^{d-4}$.}
\begin{equation}
I_{M}=-\frac{1}{4\epsilon \Omega _{d-2}}\int \sqrt{-g}F^{\mu \nu }F_{\mu \nu
}\text{ }d^{d}x.  \label{IMaxwell}
\end{equation}
Electrically charged solutions which are static and spherically symmetric
can be found through the ansatz (\ref{gspherical}), and requiring that and
the only non vanishing component of the electromagnetic field strength be
\begin{equation}
F_{0r}=-\partial _{r}A_{0}(r).  \label{espherical}
\end{equation}
The field equations for $N$, $f^{2}$and $A_{0}$ read

\begin{eqnarray}
\frac{dN}{dr} &=&0,  \nonumber \\
\frac{d}{dr}(r^{d-2}p) &=&0,  \nonumber \\
\frac{dA_{0}}{dr}+Np &=&0,  \nonumber \\
\frac{d}{dr}\left( r^{d-1}\left[ F(r)+\frac{1}{l^{2}}\right] ^{k}\right) &=&%
\frac{G_{k}}{\epsilon }r^{d-2}p^{2},  \label{eqsgm}
\end{eqnarray}
where $F(r)$\ is defined in equation (\ref{F(r)}),{\bf \ }and $p(r)$ is a
redefinition of the electric field,
\begin{equation}
p=\frac{1}{N}F_{0r}\;.  \label{p(r)}
\end{equation}
Integrating these equations yields

\begin{eqnarray}
N &=&N_{\infty }=1\;,  \nonumber \\
p(r) &=&\epsilon \frac{Q}{r^{d-2}}\;,  \nonumber \\
A_{0}(r) &=&\phi _{\infty }+\frac{\epsilon }{(d-3)}\frac{Q}{r^{d-3}}\text{\ }%
,  \nonumber \\
f^{2}(r) &=&1+\frac{r^{2}}{l^{2}}-\sigma g_{k}(r)\;,  \label{solutcharge}
\end{eqnarray}
with $\sigma =(\pm 1)^{(k+1)}$ and
\begin{equation}
g_{k}(r)=\left( \frac{2G_{k}M+\delta _{d-2k,1}}{r^{d-2k-1}}-\frac{\epsilon
G_{k}}{\left( d-3\right) }\frac{Q^{2}}{r^{2(d-k-2)}}\right) ^{\frac{1}{k}%
}\!\!.  \label{g(r)}
\end{equation}
The integration constants $M$ and $Q$ in (\ref{g(r)}) are the mass and the
electric charge of the black hole respectively, as is shown in the next
subsection.

Equations (\ref{solutcharge}) provide the electrically charged extension of
the vacuum solution (\ref{f2})\footnote{%
The expression (\ref{g(r)}) is valid for $d>3$. The three dimensional case
is discussed in Refs. \cite{BTZ,MTZ}.}. The presence of $\sigma $ in (\ref
{solutcharge}) leads to a similar picture as in the uncharged case. When $k$
is odd, there is a unique electrically charged black hole solution because $%
\sigma $ is always equal to one, but when $k$ is even, the solution with $%
\sigma =1$ represents a black hole, and the solution with $\sigma =-1$
possess a naked singularity.

Therefore, electrically charged asymptotically AdS black hole solutions are
obtained from (\ref{solutcharge}) with $\sigma =1$, whose line element read
--for $d>3$-- as
\begin{eqnarray}
ds^{2} &=&-\left( 1+\frac{r^{2}}{l^{2}}-g_{k}(r)\right) dt^{2}+  \nonumber \\
&&%
%TCIMACRO{\dfrac{dr^{2}}{1+\frac{r^{2}}{l^{2}}-g_{k}(r)}}
%BeginExpansion
{\displaystyle {dr^{2} \over 1+\frac{r^{2}}{l^{2}}-g_{k}(r)}}%
%EndExpansion
+r^{2}d\Omega _{d-2}^{2}\;,  \label{BHMQ}
\end{eqnarray}
where $g_{k}(r)$\ is given by (\ref{g(r)}).

As is naturally expected, the set of black holes described by (\ref{BHMQ}),
reduce to the $d$-dimensional Reissner-Nordstrom-AdS solution for $k=1$. The
electrically charged black hole solutions corresponding to BI and CS
theories \cite{btz} are also recovered for $d=2n$ and $d=2n-1$ respectively,
as it can be seen replacing $k=n-1$ in (\ref{BHMQ}).

For a generic value of the label $k$, in analogy with standard
Reissner-Nordstr\"{o}m-AdS geometry, the black hole solutions given by (\ref
{BHMQ}) possess in general two horizons located at the roots of{\bf \ }$%
f^{2}(r)$. They satisfy $0<r_{-}<r_{+}$ provided the mass is bounded from
below\ as $M\geq h_{k}(Q)$, where $h_{k}$ is a monotonically increasing
function of the electric charge. Both horizons merge \thinspace when the
bound is saturated, corresponding to the extreme case, that is $r_{+}=r_{-}$
for $M=h_{k}(Q)$. Solutions with $M<h_{k}(Q)$\ possess naked singularities
which should be considered unphysical. Thus, for a given electric charge,
the existence of a lower bound on $M$ is in agreement with the cosmic
censorship principle.

An important difference with the Reissner-Nordstr\"{o}m-AdS case ($k=1$)\ is
shared by all electrically charged black hole solutions with $k\neq 1$, as
can be inferred evaluating the scalar curvature for the metrics (\ref{BHMQ}%
), given by
\begin{equation}
R=\frac{1}{r^{d-2}}\frac{d^{2}}{dr^{2}}\left[ r^{d-2}\left( g_{k}(r)-\frac{%
r^{2}}{l^{2}}\right) \right] .  \label{R}
\end{equation}
For any $k\neq 1$, equation (\ref{R}) has a branch point unbounded
singularity at the zero of the function $g_{k}(r)$. This is a real timelike
singularity located at
\begin{equation}
r_{e}=\left( \frac{\epsilon }{2(d-3)}\frac{Q^{2}}{(M+\frac{1}{2G_{k}}\delta
_{d-2k,1})}\right) ^{\frac{1}{d-3}},  \label{re}
\end{equation}
which can be reached in a finite proper time. However, an external observer
is protected from it because it is surrounded by both horizons, {\em i.e.} $%
0<r_{e}<r_{-}<r_{+}$.

When $k$ is even, spacetime cannot be extended to $r<r_{e}$, because in that
case the metric (\ref{BHMQ}) would become complex. This means that the
manifold possesses a real boundary at $r=r_{e}$, and therefore, $r_{e}$ {\em %
is the\ smallest possible size} of a spherical body endowed of electric
charge $Q$ and mass $M$.

For odd values of $k\neq 1$\ there is no obstruction to define spacetime
within the region $r<r_{e}$. However, as it can be seen from (\ref{R}),
there is an additional timelike singularity located at $r=0$. In that case,
a spherical source with electric charge $Q$ and mass $M$, whose radius is
smaller than $r_{e}$ possesses an exterior geometry described by (\ref{BHMQ}%
) which cannot be empty, since it has a singularity at $r=r_{e}$. This means
that the original source generates ``a shield'', which acts as the effective
source of the external geometry. Hence again, $r_{e}$ is the{\em \ smallest
size} for the source.

This means that the presence of electric charge brings in a new length scale
into the system, except when one deals with the EH\ action. For CS theory ($%
d=2k+1$), the radius $r_{e}$ depends on the gravitational constant. However,
in the generic case, which is given by the set of theories which are neither
EH or CS, the radius $r_{e}$ depends only on intrinsic features of the
source and it is completely independent from gravity. That is, $r_{e}$ is
independent of the label $k$, the gravitational constant $G_{k}$ and the
cosmological constant -- or equivalently the AdS radius $l$ --, that is
\begin{equation}
r_{e}=\left( \frac{\epsilon }{2(d-3)}\frac{Q^{2}}{M}\right) ^{\frac{1}{d-3}},
\label{reGeneral}
\end{equation}
which has the same expression as the classical radius of the electron in $d$%
\ dimensions. It is noteworthy that $r_{e}$ is encoded in the geometry.

Remarkably, the only theory within the family discussed here, which is
unable to predict a minimum size for the source is General Relativity.

\subsection{Mass and Electric Charge from Boundary Terms}

In order to identify the integration constants appearing in the black hole
solutions\ (\ref{BHGeneral}) and (\ref{BHMQ}) with the mass and electric
charge, it is convenient to carry out the canonical analysis \cite
{Regge-Teitelboim}. The total action can be written in Hamiltonian form as

\begin{equation}
I_{T}=I_{G}+I_{M}+B\;,  \label{IHamiltonian}
\end{equation}
where $I_{G}$ and $I_{M}$ are the canonical actions for gravity and
electromagnetism, respectively

\begin{eqnarray}
I_{G} &=&\int d^{d}x(\pi ^{ij}\dot{g}_{ij}-N^{\bot }H_{G\bot }-N^{i}H_{Gi}),
\label{IGH} \\
I_{M} &=&\int d^{d}x(p^{i}\dot{A}_{i}-N^{\bot }H_{M\bot
}-N^{i}H_{Mi}-A_{0}\partial _{i}p^{i}),  \label{IMH}
\end{eqnarray}
and $B$ stands for a boundary term which is needed so that the action
attains an extremum on the classical solution. Here $H_{G\mu }$\ and $%
H_{M\mu }$\ are the Hamiltonian generators of diffeomorphisms on the
gravitational and electromagnetic phase spaces, respectively\ (see Ref. \cite
{tz}).

In case of static, spherically symmetric spacetimes, a general theorem \cite
{Palais} implies that the extremum of the action can be found through a
minisuperspace model, which is obtained replacing the {\em Ans\"{a}tze} (\ref
{gspherical}) and (\ref{espherical}) into the action, as well. Hence, one
deals with a simple one-dimensional model which allows fixing the boundary
term $B\;$as a function of the integration constants requiring the total
action (\ref{IHamiltonian}) to have an extremum on the classical solutions.
The minisuperspace action takes the form
\begin{eqnarray}
I_{T} &=&\Delta t\int \frac{N}{2}\left[ \frac{d}{dr}\left\{ \frac{r^{d-1}}{%
G_{k}}\left[ F(r)+\frac{1}{l^{2}}\right] ^{k}\right\} -\frac{1}{\epsilon }%
r^{d-2}p^{2}\right] dr  \nonumber \\
&&+\frac{1}{\epsilon }\Delta t\int A_{0}\frac{d}{dr}\left( r^{d-2}p\right)
dr+B\;,  \label{IMini}
\end{eqnarray}
where $N:=N^{\perp }(r)f^{-2}(r)$, and $p$ is a redefinition of the
canonical momentum $p^{r}$, conjugate to $A_{r}$,
\begin{equation}
p=\frac{1}{N}F_{0r}=\frac{\epsilon \Omega _{d-2}}{r^{d-2}\sqrt{\gamma }}%
p^{r},  \label{pcanonic}
\end{equation}
and $\gamma $\ is the determinant of the angular metric.

The action (\ref{IMini}) is a functional of the fields $N$, $f^{2}$, $A_{0}$%
\ and $p$, whose variation leads to a bulk term which vanishes on the field
equations (\ref{eqsgm}). Thus, the variation of the action (\ref{IMini}) on
shell is a boundary term given by

\begin{eqnarray}
\delta I_{T} &=&\Delta t\int \frac{d}{dr}\left( N\frac{r^{d-1}}{2G_{k}}%
\delta \left[ F(r)+\frac{1}{l^{2}}\right] ^{k}\right) dr  \nonumber \\
&&+\frac{1}{\epsilon }\Delta t\int \frac{d}{dr}\left( A_{0}r^{d-2}\delta
p\right) dr+\delta B\;,  \label{VarIMini}
\end{eqnarray}
which means that the action is stationary on the black hole solution provided

\begin{equation}
\delta B=-\Delta t(N_{\infty }\delta M+\phi _{\infty }\delta Q).
\label{DeltaB}
\end{equation}
Since $\delta M$ is multiplied by the proper time separation at infinity,
one identifies $M$\ and $Q$\ as the mass and the electric charge up to
additive constants. The additive constant related with the mass is called $%
C_{0}$\ and it is fixed in (\ref{C0}), requiring that the horizon shrink to
a point for\ $M\rightarrow 0$. The additive constant related with the
electric charge\ vanishes demanding that the electrically charged solution (%
\ref{BHMQ}) reduces to the uncharged one (\ref{BHGeneral}) for $Q=0$.
Therefore, the boundary term that must be added to the action is
\begin{equation}
B=-\Delta t(M+\phi _{\infty }Q)+B_{0},  \label{bt}
\end{equation}
where $N_{\infty }$ has been chosen equal to $1$, and $B_{0}$\ is an
arbitrary constant without variation.{\bf \ }This proves that the
integration constants $M$ and $Q$ appearing in the black hole metrics (\ref
{BHMQ}) and (\ref{BHGeneral}) are the mass and the electric charge
respectively.

These results are confirmed also through an alternative method which holds
for even dimensions, as is discussed in Appendix B.

\subsection{Asymptotically flat limit $(l\rightarrow \infty )$}

The black hole metrics (\ref{BHGeneral}) and (\ref{BHMQ}) tend
asymptotically to an AdS spacetime with radius $l$, whose curvature
satisfies $R^{ab}\rightarrow -l^{-2}e^{a}e^{b}$ at the boundary. Then, their
asymptotically flat limit is obtained by taking $l\rightarrow \infty $.
Thus, instead of taking the vanishing limit of the volume term $(\alpha
_{0}\rightarrow 0)$, the vanishing cosmological constant limit of the action
$I_{k}$ is obtained setting $l\rightarrow \infty $ in (\ref{Coefs}). This
procedure is consistent with taking the same limit in the field equations (%
\ref{kEinstein}) and (\ref{kTorsion}).

When $l\rightarrow \infty $ the only non-vanishing term in (\ref{Coefs}) is
the $k$th one, consequently the action is obtained from (\ref{Lovaction})
with the following choice of coefficients:

\begin{equation}
\alpha _{p}:=\tilde{c}_{p}^{k}=\frac{1}{(d-2k)}\delta _{p}^{k}\;.
\label{Coefs0}
\end{equation}
\ Therefore, replacing (\ref{Coefs0}) in (\ref{Lovaction}), a new family of
Lagrangians labeled by the integer $k\in \{1,2,...,[\frac{d-1}{2}]\}$, is
obtained. For a fixed value of $k$, the Lagrangian is given just by $L^{(k)}$
defined in (\ref{Lovlag}), so that the action reads
\begin{equation}
\tilde{I}_{k}\!=\!\frac{\kappa }{(d-2k)}\!\!\int \!\!\epsilon _{a_{1}\cdots
a_{d}}R^{a_{1}a_{2}}\!\cdot \!\cdot \!\cdot
\!R^{a_{2k-1}a_{2k}}e^{a_{2k+1}}\!\cdot \!\cdot \!\cdot \!e^{a_{d}},
\label{ActionSL}
\end{equation}
where $\kappa $ is defined in (\ref{Kappa}). The field equations coincide
with the $l\rightarrow \infty $ limit of (\ref{kEinstein}), (\ref{kTorsion}%
), which merely amounts to replacing $\bar{R}^{ab}$\ by $R^{ab}$.

Note that for $k=1$, the standard EH action without cosmological constant is
recovered, while for $k=2$ the Lagrangian is the Gauss-Bonnet density (\ref
{EGB}).

Static and spherically symmetric solutions of (\ref{ActionSL}) lead to a
similar picture as in the electrically (un)charged asymptotically AdS case:
when $k$ is odd, one obtains only one solution describing a black hole, but
for even values of $k$, two different solutions exist, one of them describes
a black hole, while the other possesses naked singularities even when the
mass bound holds.

It is simple to verify that black hole solutions of the action (\ref
{ActionSL}) correspond to the vanishing cosmological constant limit of the
solutions for pure gravity (\ref{BHGeneral}). This also holds for the
electrically charged solutions (\ref{BHMQ}).

\subsubsection{$Q=0:$}

The asymptotically flat solutions without electric charge are given by

\begin{eqnarray}
ds^{2} &=&-\left( 1-\left( \frac{2G_{k}M}{r^{d-2k-1}}\right) ^{1/k}\right)
dt^{2}+  \nonumber \\
&&%
%TCIMACRO{
%\dfrac{dr^{2}}{1-\left( \frac{2G_{k}M}{r^{d-2k-1}}\right) ^{1/k}}}
%BeginExpansion
{\displaystyle {dr^{2} \over 1-\left( \frac{2G_{k}M}{r^{d-2k-1}}\right) ^{1/k}}}%
%EndExpansion
+r^{2}d\Omega _{d-2}^{2}\;.  \label{BHSL}
\end{eqnarray}
The generic cases correspond to $d-2k-1\neq 0$, for which the metrics (\ref
{BHSL}) represent black hole solutions with an event horizon located at $%
r_{+}=(2G_{k}M)^{1/(d-2k-1)}$. As usual, their common vacuum geometry is the
flat Minkowski spacetime, and their causal structure is described through
the standard Penrose diagram of the Schwarzschild solution. In case of $k=1$
(EH), the Schwarzschild solution is recovered from (\ref{BHSL}) for $d>3$.
Exceptional cases occur when $d=2k+1$, for which the action (\ref{ActionSL})
correspond to a CS theory for the Poincar\'{e} group $ISO(d-1,1)$. Their
static, spherically symmetric solutions (\ref{BHSL})\ do not describe black
holes because they have a naked singularity at the origin. This can be
inferred from (\ref{BHCS}) because when $l\rightarrow \infty $ the horizon
recedes to infinity. For instance, in three dimensions, the solution (\ref
{BHSL}) represent a conical spacetime \cite{D-J-'tH}.

\subsubsection{$Q\neq 0$:}

The electrically charged asymptotically flat black hole solutions can be
obtained for $d>3$ from (\ref{BHMQ}) in the limit $l\rightarrow \infty $. As
for the uncharged solutions, the generic case holds for $d-2k-1\neq 0$,
whose line elements read
\begin{equation}
ds^{2}=-\left( 1-g_{k}(r)\right) dt^{2}+%
%TCIMACRO{\dfrac{dr^{2}}{1-g_{k}(r)}}
%BeginExpansion
{\displaystyle {dr^{2} \over 1-g_{k}(r)}}%
%EndExpansion
+r^{2}d\Omega _{d-2}^{2}\;,  \label{BHQSL}
\end{equation}
with $g_{k}(r)$ given by
\begin{equation}
g_{k}(r)=\left( \frac{2G_{k}M}{r^{d-2k-1}}-\frac{\epsilon G_{k}}{(d-3)}\frac{%
Q^{2}}{r^{2(d-k-2)}}\right) ^{\frac{1}{k}}\;.  \label{g(r)SL}
\end{equation}
For different generic values of the label $k$, the black hole solutions
given by (\ref{BHQSL}) resemble the Reissner-Nordstr\"{o}m one, possessing
two horizons which are found solving $g_{k}(r)=1$. As usual, these horizons
satisfy $0<r_{-}<r_{+}$ provided the mass is bounded from below by
\begin{equation}
Q^{2}\leq \frac{(d-2k-1)}{\epsilon G_{k}}\left( \frac{(d-3)G_{k}M}{d-k-2}%
\right) ^{\frac{2d-2k-4}{d-2k-1}}.  \label{Bound}
\end{equation}
The extreme case occurs when both horizons coalesce, that is
\begin{equation}
r_{+}=r_{-}=\left( \frac{(d-3)G_{k}M}{d-k-2}\right) ^{\frac{1}{d-2k-1}},
\label{Extreme}
\end{equation}
so that the bound (\ref{Bound}) is saturated.

The $d$-dimensional Reissner-Nordstrom solution is obtained from (\ref{BHQSL}%
) setting $k=1$. Equation (\ref{Bound}) reproduces the well known
four-dimensional bound given by
\begin{equation}
Q_{EH}^{2}\leq \frac{G_{{}}M^{2}}{\epsilon }\;,  \nonumber
\end{equation}
which is saturated when $r_{+}=r_{-}=G_{k}M$, as can be seen from (\ref
{Extreme}) for $d=4$ and $k=1$.

A further example corresponds to the electrically charged black hole in the
vanishing cosmological constant limit of the BI action. The bound and the
extreme radius are obtained in that case from (\ref{Bound}) and (\ref
{Extreme}) for $d=2n$ and $k=n-1$:

\begin{eqnarray}
Q_{BI}^{2} &\leq &\frac{1}{\epsilon G_{n-1}}\left[ \frac{(2n-3)G_{n-1}M}{n-1}%
\right] ^{2(n-1)}  \nonumber \\
r_{+} &=&r_{-}=\frac{(2n-3)G_{n-1}M}{n-1}.
\end{eqnarray}

The full set of asymptotically flat electrically charged black hole
solutions (\ref{BHQSL}) share a common feature with its asymptotically AdS
counterparts given by (\ref{BHMQ}) in the generic case $(d-2k-1\neq 0)$.
That is the existence of a timelike singularity for $k\neq 1$ located at the
zero of $g_{k}(r)$ in (\ref{g(r)SL}) given by
\begin{equation}
r_{e}=\left( \frac{\epsilon }{2(d-3)}\frac{Q^{2}}{M)}\right) ^{\frac{1}{d-3}%
},
\end{equation}
which satisfies $0<r_{e}<r_{-}<r_{+}$ and is again interpreted as {\em the\
smallest possible size} of a spherical body with electric charge $Q$ and
mass $M$. Then one concludes that this feature is absent only when one deals
with the EH action with or without cosmological constant.

\section{Thermodynamics}

\subsection{Temperature}

As usual, we define the black hole temperature by the condition that in the
Euclidean sector, the solution be well defined (smooth) at the horizon. This
means that the Euclidean time is a periodic coordinate with period
\begin{equation}
\tau =4\pi \left( \left. \frac{df^{2}}{dr}\right| _{r_{+}}\right) ^{-1}\;,
\label{period}
\end{equation}
which is identified with $\beta =\frac{1}{\kappa _{B}T}$, where $\kappa _{B}$
is the Boltzmann constant. Thus, the Hawking temperature is given by

\begin{equation}
T=\frac{1}{4\pi \kappa _{B}}\left. \frac{df^{2}}{dr}\right| _{r_{+}}\;.
\label{temperature}
\end{equation}

For the electrically uncharged cases, the black hole temperature for the set
of metrics (\ref{BHGeneral}) is

\begin{equation}
T=\frac{1}{4\pi \kappa _{B}k}\left( (d-1)\frac{r_{+}}{l^{2}}+\frac{(d-2k-1)}{%
r_{+}}\right) .  \label{temp}
\end{equation}
For all $k$ such that $d-2k-1\neq 0$,\ the function $T(r_{+})$\ exhibits the
same behavior as the standard Schwarzschild-AdS black hole (which is
obtained for $k{\bf =}1$), that is: the temperature diverges at $r_{+}=0$.
It\ has a minimum at $r_{c}$ given by

\begin{equation}
r_{c}=l\sqrt{\frac{d-2k-1}{d-1}},  \label{rc}
\end{equation}
and grows linearly for large $r_{+}$. Considering $k=n-1$, formula (\ref
{temp}) reproduces the known results for BI ($d=2n$) and CS ($d=2n-1$) black
holes \cite{btz}.\ The temperature (\ref{temp}) reaches an absolute minimum
at $r_{c}$ equal to
\begin{equation}
T_{c}=\frac{\sqrt{\left( d-2k-1\right) \left( d-1\right) }}{2\pi \kappa
_{B}kl},
\end{equation}
provided the existence of a nonvanishing cosmological constant ($l\neq
\infty $).

In case of CS theory, that is when $d-2k-1=0$, $T(r_{+})$\ is not
divergent at all, its absolute minimum is at $r_{c}=0$ and
$T_{c}=0$. Thus, CS black holes are the only exceptional cases
among all the possibilities considered here. Both, CS and generic
cases are depicted in Figure $1$.

{\begin{figure}
\begin{center}
\leavevmode \epsfxsize=3in \epsfbox{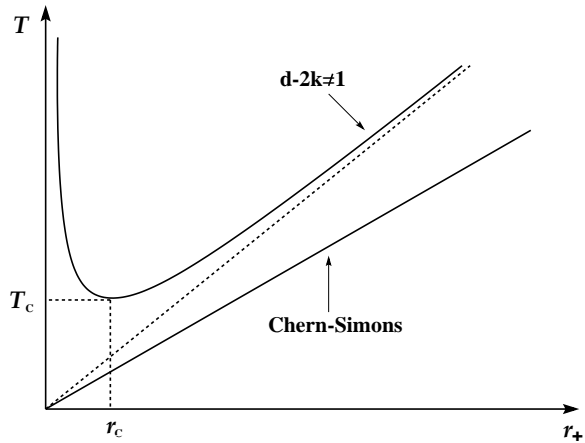}\label{fig1}
\caption{The black hole temperature is plotted as a function of
the horizon radius $r_{+}$. For $d-2k\neq 1$\ the temperature
reaches an absolute minimum $T_{c}$\ at $r_{+}=r_{c}$.}
\end{center}
\end{figure}

\subsection{Specific Heat and Thermal Equilibrium}

As seen in Section III.A{\bf ,} the black hole mass is a monotonically
increasing function of $r_{+}$, therefore\ the behavior of $T(M)$\ is
qualitatively similar to that of $T(r_{+})$.

Using (\ref{temp}) and (\ref{mass(r)}), the specific heat $C_{k}=\frac{%
\partial M}{\partial T}$, can be expressed as a function of $r_{+}$,
\begin{equation}
C_{k}=k\frac{2\pi \kappa _{B}}{G_{k}}r_{+}^{d-2k}\left( \frac{%
r_{+}^{2}+r_{c}^{2}}{r_{+}^{2}-r_{c}^{2}}\right) \left( 1+\frac{r_{+}^{2}}{%
l^{2}}\right) ^{k-1},  \label{capcal}
\end{equation}
In case of $d-2k-1\neq 0$, the specific heat (\ref{capcal}) possesses an
unbounded discontinuity at $r_{+}=r_{c}$ (see Figure $1$), signaling a phase
transition. The specific heat $C$ is positive for $r_{+}>r_{c}$, and has the
opposite sign for $r_{+}<r_{c}$.

Again, the CS case is exceptional. Setting $d=2n-1$\ and $k=n-1$ in (\ref
{capcal}), the specific heat is found as
\begin{equation}
C_{CS}=\left( n-1\right) \frac{2\pi \kappa _{B}}{G_{n-1}}r_{+}\left( 1+\frac{%
r_{+}^{2}}{l^{2}}\right) ^{n-2},  \label{capcalcs}
\end{equation}
which is a continuous monotonically increasing positive function of $r_{+}$
and does not diverge for any finite value of $r_{+}$ \cite{Muniain-Piriz}.

\begin{figure}
\begin{center}
\leavevmode \epsfxsize=3in
\epsfbox{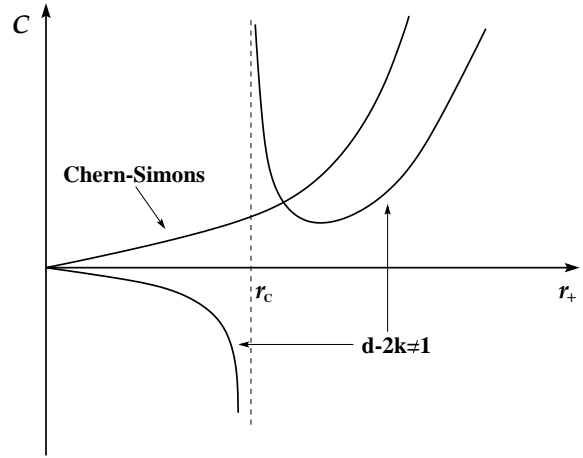}\label{fig2}\caption{The specific heat $C_{k}$\
is plotted as a function of the horizon
radius. For a generic theory, $d-2k\neq 1$, $C_{k}$\ has a simple pole at $%
r_{+}=r_{c}$. For the exceptional case, $d=2k+1$\ (CS), the
specific heat is a continuous, monotonically increasing, positive
function of $r_{+}$.}
\end{center}
\end{figure}

The presence of a negative cosmological constant makes it possible for the
family of black hole solutions (\ref{BHGeneral}) to reach thermal
equilibrium, as is possible for the Schwarzschild-AdS$_{4}$ spacetime \cite
{Hawking-Page} and for the three-dimensional black hole. Let us assume that
any black hole described by (\ref{BHGeneral}) is immersed in a thermal bath
of temperature $T_{0}>T_{c}$. If $d-2k-1\neq 0$, the thermal behavior splits
in two branches: for $r_{+}<r_{c}$, the specific heat is negative and
therefore black hole state is driven away from that with temperature $T_{0}$%
; for $r_{+}>r_{c}$, the black hole state is attracted towards the
equilibrium configuration at temperature $T_{0}$ (see Figure $3$). Thus, the
temperature $T_{0}$ corresponds to two equilibrium states of radii $r_{u}$
(unstable) and $r_{s}$ (locally stable), with $r_{u}<r_{c}<r_{s}$.{\bf \ }%
Neglecting quantum tunneling processes, there are two possible scenarios: if
the initial black hole state has $r_{+}<r_{u}$, the black hole cannot reach
the equilibrium because it evaporates until its final stage. Otherwise, for $%
r_{+}>r_{u}$, the black hole evolves towards an equilibrium configuration at
$r_{+}=r_{s}$.
\begin{figure}
\begin{center}
\leavevmode \epsfxsize=3in
\epsfbox{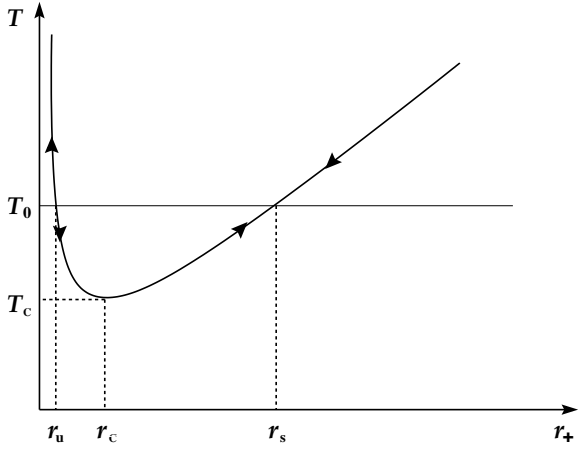}\label{fig3}\caption{In the generic case,
$d-2k\neq 1$,\ the black hole can reach thermal equilibrium with
a bath of temperature higher than $T_{c},$\ provided the horizon
radius satisfies $r_{+}>r_{u}$.}
\end{center}
\end{figure}

If the heat bath has temperature below $T_{c}$, the black hole cannot reach
a stable equilibrium state and must evaporate, as depicted in Figure $4$.

\begin{figure}
\begin{center}
\leavevmode \epsfxsize=3in
\epsfbox{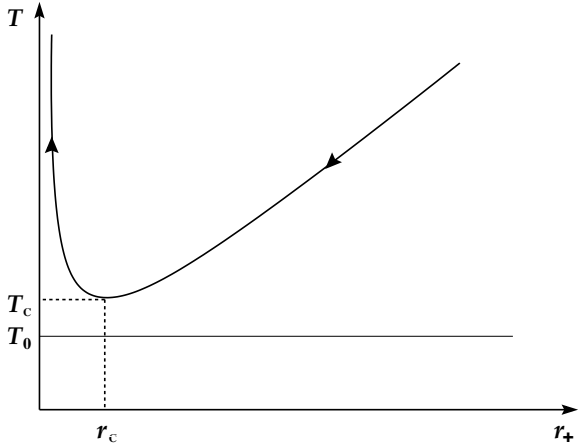}\label{fig4}\caption{In the generic case,
$d-2k\neq 1,$\ the black hole cannot reach thermal equilibrium
with a bath of temperature lower than $T_{c}$.}
\end{center}
\end{figure}

None of the above arguments hold for the Chern-Simons case. When $d-2k=1$,
the specific heat (\ref{capcalcs}) is always positive, therefore the
equilibrium configuration is always reached, independently from the initial
black hole state and for any finite temperature of the heat bath.

\subsection{Entropy}

It is well known that the partition function which describes the black hole
thermodynamics is obtained through the Euclidean path integral in the saddle
point approximation around the black hole solution \cite{Gibbons-Hawking}.
That is,

\[
Z\approx e^{-I_{E}},
\]
which means that the Euclidean action evaluated on the black hole
configuration is identified with $\beta $ times the free energy of the system

\begin{equation}
I_{E}=\beta M-\frac{S}{\kappa _{B}}+\beta \sum\limits_{i}\mu _{i}Q_{i}.
\label{IE}
\end{equation}
where the $\mu _{i}$'s are the chemical potentials corresponding to the
charges $Q_{i}$. The Euclidean minisuperspace action is given by the
Wick-rotated form of (\ref{IMini}), that is

\begin{eqnarray}
I_{E} &=&-\beta \int\limits_{r_{+}}^{\infty }\frac{N}{2}\left[ \frac{d}{dr}%
\left\{ \frac{r^{d-1}}{G_{k}}\left[ F(r)+\frac{1}{l^{2}}\right] ^{k}\right\}
-\frac{1}{\epsilon }r^{d-2}p^{2}\right] dr+  \nonumber \\
&&-\frac{1}{\epsilon }\beta \int\limits_{r_{+}}^{\infty }A_{0}\frac{d}{dr}%
\left( r^{d-2}p\right) dr+B_{E},  \label{Euclidaction}
\end{eqnarray}

In what follows we shall consider the electrically uncharged cases only. The
bulk part of the Euclidean action is a linear combination of the constrains
and therefore, its on-shell value is given by the boundary term $B_{E}$.
This boundary piece is determined by the requirement that $I_{E}$ be
stationary on the black hole geometry. Varying (\ref{Euclidaction}) leads to
\begin{equation}
\delta I_{E}=-\frac{\beta N_{\infty }}{2G_{k}}\int\limits_{r_{+}}^{\infty }%
\frac{d}{dr}\left\{ r^{d-1}\delta \left[ F(r)+\frac{1}{l^{2}}\right]
^{k}\right\} dr+\delta B_{E},  \label{VarIE}
\end{equation}
on shell. From this expression, one finds
\[
\delta B_{E}=\beta \delta M-\frac{2\pi k}{G_{k}}r_{+}^{d-2k-1}\left( 1+\frac{%
r_{+}^{2}}{l^{2}}\right) ^{k-1}\delta r_{+},
\]
where $N_{\infty }$ has been set equal to one and we have used $\left. \frac{%
df^{2}}{dr}\right| _{r_{+}}=4\pi \beta ^{-1}$. From (\ref{IE}) one identifies

\begin{equation}
\delta S=k\frac{2\pi \kappa _{B}}{G_{k}}r_{+}^{d-2k-1}\left( 1+\frac{%
r_{+}^{2}}{l^{2}}\right) ^{k-1}\delta r_{+},
\end{equation}
which is integrated into
\begin{equation}
S_{k}=k\frac{2\pi \kappa _{B}}{G_{k}}%
%TCIMACRO{\dint }
%BeginExpansion
\displaystyle \int %
%EndExpansion
\limits_{0}^{r_{+}}r^{(d-2k-1)}\left( 1+\frac{r^{2}}{l^{2}}\right) ^{k-1}dr.
\label{Entropy}
\end{equation}
This is a monotonically increasing function of $r_{+}$, in agreement with
the second law of thermodynamics. In (\ref{Entropy}) the lower limit in the
integral has been fixed by the condition $S_{k}(r_{+}=0)=0$ for the whole
set of black holes given by (\ref{BHGeneral}).

For the EH action (that is for $k=1$), expression (\ref{Entropy}) readily
reproduces, for the Schwarzschild-AdS solution
\[
S_{EH}=\frac{2\pi \kappa _{B}}{(d-2)G}r_{+}^{d-2},
\]
which in standard units \cite{GTilde} is the celebrated ``area law''
\[
S_{EH}=\frac{\kappa _{B}}{\tilde{G}}\frac{A}{4}.
\]

For $k=\left[ \frac{d-1}{2}\right] $ (BI and CS), formula (\ref{Entropy})
reduces to the known results \cite{btz}. The theory described by $I_{2}$ in (%
\ref{GB+}) is an intrinsically higher dimensional one, and the corresponding
black hole entropy is given by
\begin{equation}
S_{2}=\frac{4\pi \kappa _{B}}{G_{{\bf 2}}}r_{+}^{d-4}\left[ \frac{1}{(d-4)}+%
\frac{r_{+}^{2}}{(d-2)l^{2}}\right] .  \label{SGB}
\end{equation}

Hence, the area law is a peculiarity of the Einstein-Hilbert theory ($k=1$),
while for $k\neq 1$ the entropy (\ref{Entropy}) becomes proportional to the
area in the large $r_{+}$ limit, that is
\begin{equation}
S_{k}\approx k\frac{2\pi \kappa _{B}}{(d-2)G_{k}l^{2(k-1)}}r_{+}^{d-2}=k%
\frac{G}{G_{k}l^{2(k-1)}}S_{EH}\;,  \label{LargeEntropy}
\end{equation}
with $r_{+}>>l$.

\subsection{Asymptotically flat limit}

In the limit $l\rightarrow \infty $,\ the geometry of the uncharged black
hole is given by (\ref{BHSL}) whose corresponding temperature is
\begin{equation}
T^{0}=\frac{1}{4\pi \kappa _{B}k}\frac{(d-2k-1)}{r_{+}}.  \label{Tlim}
\end{equation}
This gives a vanishing value for CS theory $(d-2k-1=0)$, which is consistent
with the fact that in that case, the geometry possesses a singularity which
is not surrounded by a horizon in the limit $l\rightarrow \infty $, so that
no temperature can be associated with it. For all the other cases $%
(d-2k-1\neq 0)$, the horizon is located at{\bf \ }$%
r_{+}=(2G_{k}M)^{1/(d-2k-1)}$, so that the black hole temperature (\ref{Tlim}%
) is a monotonically decreasing function of the mass. Therefore, thermal
equilibrium can never be reached, consistently with the fact that the
specific heat is always negative
\begin{equation}
C^{0}=-k\frac{2\pi \kappa _{B}}{G_{k}}r_{+}^{d-2k}.  \label{Clim}
\end{equation}
The entropy is also an increasing function of $r_{+}$,
\begin{equation}
S_{k}^{0}=k\frac{2\pi \kappa _{B}}{G_{k}}\frac{r_{+}^{(d-2k)}}{(d-2k)},
\label{SL0}
\end{equation}
which is consistent with the second law of thermodynamics. Note that formula
(\ref{SL0}) is proportional to the area of the horizon only for $k=1$ (EH).
Thus, in the $l\rightarrow \infty $ limit, the area law cannot be recovered
even as an approximation in the cases with $k\neq 1$.

\subsection{Canonical Ensemble}

In four dimensions, Hawking and Page have shown that in the presence of a
negative cosmological constant, the partition function in the canonical
ensemble is well defined, unlike in case of a vanishing $\Lambda $ \cite
{Hawking-Page}. The same argument can be extended for higher dimensions for
the whole set of theories (\ref{Ik}) labelled by $k$.

The partition function in the canonical ensemble reads
\begin{equation}
Z(\beta )=\int_{0}^{\infty }e^{-\beta M}\rho (M)dM\;,  \label{Z}
\end{equation}
where $\rho (M)=\exp \left( \frac{S_{k}}{\kappa _{B}}\right) $ is the
density of states as a function of the energy. The convergence of this
integral depends on the asymptotic behavior of $S_{k}$ for large $M$,
\[
S_{k}\approx a_{d,k}M^{\left( \frac{d-2}{d-1}\right) }\;,
\]
where $a_{d,k}$ is a positive constant. Thus, the integrand of (\ref{Z})
goes as $\exp (-\beta M+\kappa _{B}^{-1}a_{d,k}M^{\left( \frac{d-2}{d-1}%
\right) })$ and therefore the partition function converges.

This argument breaks down in the $l\rightarrow \infty $ limit: in that case,
the entropy is
\[
S_{k}^{0}=a_{d,k}^{0}M^{\left( \frac{d-2k}{d-2k-1}\right) }\;,
\]
with $a_{d,k}^{0}$ a different positive constant, which yields a divergent
partition function.

The lesson one can draw from this exercise is that the presence of a
negative cosmological constant is sufficient to render the canonical
ensemble well defined for all the theories described here.

\section{Summary and Discussion}

\subsection{Theories described by the action $I_{k}$}

We have examined a family of gravitation theories in dimension $d$, whose
common feature is to possess vacuum solutions with maximal symmetry. This
means that the theories --described by the action $I_{k}$-- have a unique
cosmological constant. For a given $d$ there exist $[\frac{d-1}{2}]$
different theories labeled by the integer $k$, which is the highest power of
curvature in the Lagrangian. For $k=1$, the EH action is recovered, while
for the largest value of $k$, that is $k=[\frac{d-1}{2}]$, BI and CS\
theories are obtained. These three cases exhaust the different possibilities
up to six dimensions, and new interesting cases arise for $d\geq 7$. For
instance, the case with $k=2$, which is described by the action (\ref{GB+}),
exists only for $d>4$: In five dimensions this theory is equivalent to CS,
for $d=6$ it is equivalent to BI, and for $d=7$ and up, it defines a new
class of theories.

\subsection{Special cases selected from cosmic censorship}

A first distinction between the different theories mentioned
above comes from the study of their spherically symmetric, static
solutions. It is found that for odd $k$, physical black holes
satisfying the cosmic censorship criterion exist. For even $k$,
however, both physical black holes and solutions with naked
singularities with positive mass exist. This already casts doubt
on the soundness of this subset of theories. Moreover, the
absence of a cosmic censorship principle would be in conflict
with the existence of a positive energy theorem obtained from
supersymmetry. This means that the supersymmetric extensions of
the theories considered here can be expected to be very different
for odd and even $k$. In fact, as it has been shown in
\cite{trz,trz2}, CS\ theories with even $k$ --defined for
$d=5,9,... $--, have a supersymmetric extension based on
superunitary groups, whereas for odd $k$ $(d=3,7,11,...)$ the
corresponding supergravities are based on the orthosymplectic
groups.

The different theories considered here are summarized in the scheme shown in
Fig. 5.

\begin{figure}
\begin{center}
\leavevmode \epsfxsize=3in
\epsfbox{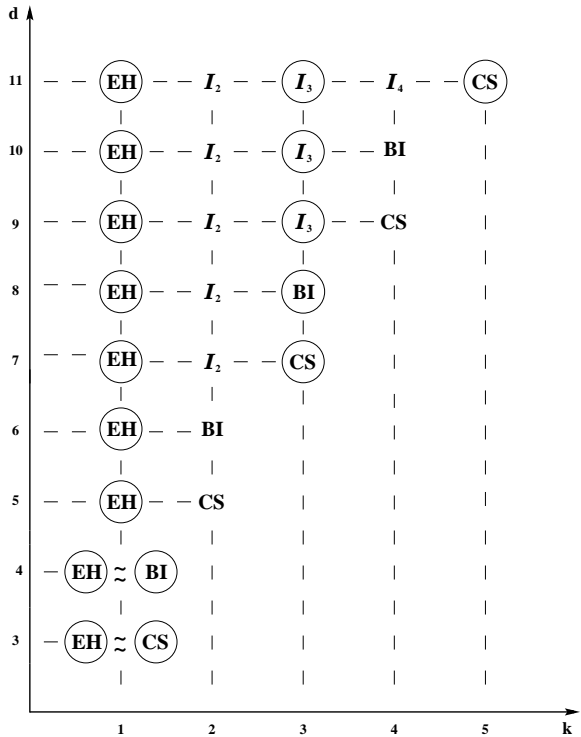}\label{fig5}\caption{Black Hole Scan: Summary
of all theories described by $I_{k}$ up to eleven dimensions. The
integer $k=1,...,[\frac{d-1}{2}]$ represents the highest power of
curvature in the action. The columns with odd $k$ are singled out
by cosmic censorship. The supersymmetric extensions of EH and CS
theories are known. The supergravities for the remaining
$I_{k}$'s are unknown.}
\end{center}
\end{figure}

Here we have highlighted the odd $k$ columns as they would represent better
candidates for physical theories based on the criterion of cosmic censorship
versus supersymmetry.

Note that CS theories are the representatives of the lowest possible
dimension for a given $k$. Moreover, CS\ gravity theories exhibit local AdS
symmetry whereas all other gravitation theories of the same dimension only
have local Lorentz invariance (see Appendix A).

Over the years, 11-dimensional spacetime has been believed to be
the arena for the ultimate unified theory. From the present
analysis, it follows that in $d=11$, the cases $k=1,3,5$ are of
special interest. The supersymmetric extension for $k=1$ is the
famous Cremmer-Julia-Scherk supergravity \cite {C-J-S}, which
only exists if the cosmological constant vanishes \cite
{B-D-H-S}. The supersymmetric extension for $k=5$\ with a finite
$\Lambda $\ is also known \cite{trz',trz}, whose vanishing
cosmological constant version is described in \cite{BTrZ}. The
corresponding supersymmetric extension of the gravity theory with
$k=3$ is an open problem.

\subsection{Black Holes}

For all dimensions and for any $k$, there exist well behaved black hole
solutions, in the sense that the singularities are hidden by an event
horizon. For $d-2k\neq 1$, the causal structure of these black holes is the
same as that of Schwarzschild-AdS and Reissner-Nordstr\"{o}m-AdS spacetimes.
However, this set of black holes differs from standard $d$-dimensional
Schwarzschild and Reissner-Nordstr\"{o}m solutions in that their asymptotic
behavior, with respect to the vacuum, is given by $g_{00}-\bar{g}%
_{00}\approx r^{-\left( \frac{d-2k-1}{k}\right) }$. Again, the CS case
stands separate from the rest, in that the causal structure of the vacuum is
the same as that of $2+1$ dimensions, and analogously, there is a mass gap
between the $M=0$ black hole and AdS spacetime $\left( M=-\frac{1}{2G_{n-1}}%
\right) $. Furthermore, in the vanishing cosmological constant limit, the CS
theory supports no static, spherically symmetric black holes.

In the electrically charged case, the black holes for $k\neq 1$ predict a
minimum size for a physical source. It is noteworthy that the geometry
encodes this restriction for all cases, except for the EH action.

\subsection{Thermodynamics}

The presence of a negative cosmological constant for the entire set of
theories described by the action $I_{k}$ makes it possible for black holes
to reach thermal equilibrium with a heat bath. The AdS radius $l$ acts as a
regulator allowing the canonical ensemble to be well defined, unlike the
case of zero cosmological constant. The black hole entropy obeys the area
law only in the case $k=1$. For other values of $k$, the entropy respects
the second law of thermodynamics, because $\frac{dS}{dr_{+}}>0$, but the
area law is recovered only in the limit $\frac{r_{+}}{l}\rightarrow \infty $.

In the limit $\Lambda \rightarrow 0$, the area law never holds, except for $%
k=1$. In that limit, the temperature has no minimum and consequently the
thermodynamic equilibrium cannot be reached.

The thermodynamic behavior is qualitatively the same as the Schwarzschild-AdS%
$_{4}$ black hole in the generic cases $d-2k\neq 1$. On the other hand,
Chern-Simons black holes for odd dimensions behave like the $d=3$ case.

In the generic cases, black holes have a minimum temperature $T_{c}$ at $%
r_{+}=r_{c}=l\sqrt{\frac{d-2k-1}{d-1}}$, so that --as is depicted in Figure $%
3$-- those whose horizon radius exceed the unstable equilibrium position $%
r_{u}$\ can reach equilibrium with a heat bath of temperature higher than $%
T_{c}$. If the heat bath has a temperature below $T_{c}$, or $r_{+}<r_{u}$,
the black holes evaporate.

In the CS case, the temperature grows linearly with $r_{+}$, hence there is
no critical temperature and the thermal equilibrium is always attained.

In an equilibrium configuration, the free energy $F=M-TS$ can be expressed
as a function of $r_{+}$. For fixed $k$ the behavior of $F$ can be found
from (\ref{mass(r)}), (\ref{temp}) and (\ref{Entropy}) as

\begin{mathletters}
\begin{eqnarray}
F(r_{+} &\rightarrow &0)\sim \frac{r_{+}^{d-2k-1}}{2(d-2k)G_{k}}\;,
\label{I(0)} \\
F(r_{+} &\rightarrow &\infty )\sim -\frac{r_{+}^{d-1}}{2(d-2)G_{k}l^{2k}}\;.
\label{I(infty)}
\end{eqnarray}
\ This change in sign has been interpreted as an indication that, for small $%
r_{+}$ the black hole would be unstable for decay into AdS spacetime, while
for large $r_{+}$ the black hole would be stable \cite{Witten-Thermal}. This
suggests that a phase transition would occur at $F(r_{+})=0$. This
conclusion, however contradicts the fact that the phase transition actually
occurs at the critical value $r_{c}$, where the specific heat $C$ changes
sign, and which does not coincide with the zero of $F(r_{+})$. In
particular, considering the EH action $(k=1)$, the change of sign in $F$
occurs at $r_{+}=l$ while $r_{c}=l\sqrt{\frac{d-3}{d-1}}<l$. Moreover, for
the CS case, $d-2k=1$, there is no phase transition at all, although $F$
still has a change in sign. The source of the disagreement lies in that the
canonical ensemble is defined keeping $T$ fixed, while the limits in (\ref
{I(0)}) and (\ref{I(infty)}) do not respect this condition.

>From all the evidence presented here, it is apparent that CS theories form
an exceptional class: They are genuine gauge theories whose supersymmetric
extension is known; their black hole spectrum has a mass gap separating it
from AdS spacetime, and these black holes possess remarkable thermodynamical
properties. CS black holes can reach thermal equilibrium with a heat bath at
any temperature, and the positivity of the specific heat guarantees their
stability under thermal fluctuations.

In contrast with the generic case, a small CS black hole is stable against
decay by Hawking radiation. This suggests that, as in the three dimensional
case, CS\ (super)gravities could have a well defined quantum theory.

\section{Acknowledgments}

The authors are grateful to R. Aros, M. Ba\~{n}ados, M. Contreras, M.
Henneaux, C. Mart\'{\i }nez, F. M\'{e}ndez, R. Olea, M. Plyushchay, J.
Saavedra and C. Teitelboim for many enlightening discussions and helpful
comments. This work was supported in part through grants 1990189, 1980788
from FONDECYT, and by the ``Actions de Recherche Concert{\'{e}}es'' of the
``Direction de la Recherche Scientifique - Communaut{\'{e}} Fran{\c{c}}aise
de Belgique'', by IISN - Belgium (convention 4.4505.86).\ The institutional
support of Fuerza A\'{e}rea de Chile, I.\ Municipalidad de Las Condes, and a
group of Chilean companies (AFP Provida,  CODELCO, Empresas CMPC, and Telef%
\'{o}nica del Sur) is also recognized. CECS is a Millenium Science
Institute. J. Z. wishes to thank the organizers of the $1999$ ICTP Summer
Workshop on Black Hole Physics for hospitality in Trieste. J. C. and J.Z
thank the organizers of the {\em V La Hechicera School}, M\'{e}rida.

\section{Appendix}

\subsection{CS \& BI Theories}

Requiring that the integrability conditions of equation (\ref{E-L}) do not
impose further algebraic constraints on the curvature or the torsion beyond
Eq. (\ref{Tor}) implies that the coefficients $\alpha _{p}$'s in Eq. (\ref
{Lovaction}) satisfy a recursive equation, whose solution fixes them in
terms of the gravitational and cosmological constants \cite{HDG}. An
equivalent way to express this is that the $\alpha _{p}$'s become fixed as
in equation (\ref{Coefs}) with $k=[\frac{d-1}{2}]$, just requiring the
existence of a sector in the theory with propagating torsion. Thus, in $d=2n$
dimensions, the Lagrangian reads
\end{mathletters}
\begin{equation}
L=\frac{\kappa l^{2}}{2n}\epsilon _{a_{1}\cdots a_{d}}\bar{R}%
^{a_{1}a_{2}}\cdots \bar{R}^{a_{d-1}a_{d}},  \eqnum{A1}  \label{BI}
\end{equation}
where\footnote{%
\medskip A positive cosmological constant is obtained making $%
l^{2}\rightarrow -l^{2}$.} $\bar{R}^{ab}:=R^{ab}+\frac{1}{l^{2}}e^{a}e^{b}$.

The expression (\ref{BI}) is proportional to the Pfaffian of the $2$-form $%
\bar{R}^{ab}$ and, in this sense, it has a Born-Infeld-like form \cite{JJG}:
\begin{equation}
L=2^{n-1}(n-1)!\kappa l^{2}\sqrt{\det \left( R^{ab}+\frac{1}{l^{2}}%
e^{a}e^{b}\right) }.  \eqnum{A2}  \label{BI'}
\end{equation}

For $d=2n-1$ dimensions, the Lagrangian is given by the Euler-Chern-Simons
form for the AdS group, whose exterior derivative is proportional to the
Euler density in $2n$ dimensions,

\begin{eqnarray}
dL_{G\;2n-1}^{AdS} &=&\frac{\kappa l}{2n}\epsilon _{A_{1}\cdots A_{2n}}\bar{R%
}^{A_{1}A_{2}}\cdots \bar{R}^{A_{2n-1}A_{2n}}  \nonumber \\
&=&\bar{\kappa}{\cal E}_{2n},  \eqnum{A3}  \label{dCS}
\end{eqnarray}
where $\bar{R}^{AB}$ stands for the AdS curvature. This Lagrangian was
discussed in \cite{chamslett} and also in \cite{btz} for torsion-free
manifolds.

Additional terms which depend explicitly on the torsion are required by
local supersymmetry \cite{trz,trz'} and they can be consistently added to
the Lagrangian only for $d=4m-1$ \cite{HDG}.

These torsional Lagrangians are odd under parity and are obtained from the
Chern characters associated with the AdS curvature in $4m$ dimensions.
Furthermore, the coefficients in front of the different terms in these
torsional Lagrangians are necessarily quantized. The odd dimensional action,
with or without torsional terms, has a larger local symmetry given by $%
SO(d-1,2)$, so that beyond standard local Lorentz symmetry ($\delta
e^{a}=\lambda _{\;b}^{a}e^{b}$ and $\delta \omega ^{ab}=-D\lambda ^{ab}$),
these theories are invariant also under local ``AdS-translations:''
\begin{eqnarray}
\delta e^{a} &=&-D\lambda ^{a}  \nonumber \\
\delta \omega ^{ab} &=&\frac{1}{l^{2}}(\lambda ^{a}e^{b}-\lambda ^{b}e^{a}).
\eqnum{A4}  \label{transfAdS}
\end{eqnarray}

\subsection{Conserved Charges from a Background-Independent Surface Integral}

If one deals with more general solutions possessing different isometries,
the identification of the integration constants with the conserved charges
through the minisuperspace trick does not work, because in general the
reduced action does not lead to the true extremum of the original action.
The Hamiltonian method provides a way to express the mass as a surface
integral \cite{Regge-Teitelboim}. However, this procedure requires the
invertibility of the symplectic matrix associated with the action $I_{k}$.
This is impossible to perform globally in phase space, because there are
field configurations for which the symplectic form degenerates. Therefore,
no general formula could be found for an arbitrary field configuration.

A way to circumvent this problem is carried out in $d=2n$ following a
recently proposed method \cite{ACOTZ3+1,ACOTZ2n} which is appropriate to
deal with asymptotically AdS spacetimes.

Consider the action $I_{k}$ defined in (\ref{Ik}). In first order formalism,
the existence of an extremum of $I_{k}$ for asymptotically locally AdS
spacetimes fixes\ the boundary term that must be added to the action as
being proportional to the Euler density multiplied by a fixed weight factor.
Hence, in order to cancel the boundary term coming from the variation of $%
I_{k}$, the total action including the boundary term --up to a constant--,
is given by

\begin{equation}
I_{T}=I_{k}+\kappa \alpha _{n}\int {\cal E}_{2n}\;,  \eqnum{B1}
\label{IkEuler}
\end{equation}
with
\begin{equation}
\alpha _{n}=c_{n}^{k}:=\frac{(-1)^{n+k+1}l^{2(n-k)}}{2n\left(
%TCIMACRO{\QATOP{n-1}{k} }
%BeginExpansion
{n-1 \atop k}%
%EndExpansion
\right) }.  \eqnum{B2}  \label{AlphaN}
\end{equation}
The total action $I_{T}$ is invariant under diffeomorphisms by construction,
because $I_{k}$ is written in terms of differential forms. Thus, Noether's
theorem provides a conserved current $(d*J=0)$ associated with this
invariance, which can be locally written as $*J=dQ$. Assuming the topology
of the manifold to be of the form ${\cal M}=R\times \Sigma $, this procedure
yields a regularized and background-independent expression for the conserved
charges associated with a Killing vector $\xi $, which is globally defined
on the boundary of the spatial section $\partial \Sigma $. The surface
integral reads

\begin{equation}
Q(\xi )=\int\limits_{\partial \Sigma }\xi ^{\mu }\omega _{\mu }^{ab}{\cal T}%
_{ab}\;,  \eqnum{B3}  \label{Qchi}
\end{equation}
where, ${\cal T}_{ab}$ is the variation of the total Lagrangian with respect
to the curvature
\begin{equation}
{\cal T}_{ab}:=\frac{\delta L_{T}}{\delta R^{ab}}=\sum_{p=1}^{n}c_{p}^{k}p%
{\cal T}_{ab}^{p}\;,  \eqnum{B4}  \label{Phi}
\end{equation}
with
\begin{equation}
{\cal T}_{ab}^{p}=\kappa \epsilon _{aba_{3}\cdots a_{d}}R^{a_{3}a_{4}}\cdots
R^{a_{2p-1}a_{2p}}e^{a_{2p+1}}\cdots e^{a_{d}}\;,  \eqnum{B5}  \label{Taup}
\end{equation}
and where the coefficients $c_{p}^{k}$ are defined through equations (\ref
{Coefs}) and (\ref{AlphaN}).

The mass is obtained from (\ref{Qchi}) when $\xi =\partial _{t}$, without
making further assumptions about the matching with a background geometry nor
with its topology.

One way to check this result is evaluating the mass for the black hole
metrics (\ref{BHGeneral}), which leads to the expected result
\begin{equation}
Q(\partial _{t})=M.  \eqnum{B6}
\end{equation}
It is a simple exercise to check that formula (\ref{Qchi}) vanishes when
evaluated on any constant curvature spacetime -- satisfying $\bar{R}%
^{ab}=R^{ab}+l^{-2}e^{a}e^{b}=0$ -- which admits at least one Killing
vector. This means that spaces which are locally AdS have vanishing Noether
charges for the whole set of theories defined by $I_{k}$ in even dimensions.
These spaces in general possess non-trivial topologies and could be regarded
as different possible vacua. Hence one can find massive solutions which
correspond to excitation of the corresponding vacuum in the same topological
sector.


\newpage

\begin{references}
\bibitem{Hawking-Penrose}  S. W. Hawking and R. Penrose, Proc. Roy. Soc.
Lond. {\bf A314 }(1970){\bf \ }529.

\bibitem{Witten-Energy}  E. Witten, Comm. Math. Phys.{\bf \ 80}, (1981) 381.

\bibitem{GTilde}  It is the standard practice to fix the coefficient in
front of the EH action as $-(16\pi \tilde{G})^{-1}$ for all spacetime
dimensions (see e.g. Refs. \cite{BK,Witten-Thermal}). We use a different
convention which is useful because it simplifies the expression for the
black hole metrics in higher dimensions, however it gives an slightly
unusual factor for the entropy. This choice of units is related with the
standard one through $G=\frac{8\pi }{(d-2)\Omega _{d-2}}\tilde{G}$, where $%
\Omega _{p}=\frac{2\pi ^{\frac{p+1}{2}}}{\Gamma (\frac{p+1}{2})}$ stands for
the volume of $S^{p}$.

\bibitem{BK}  V. Balasubramanian and P. Kraus, Comm. Math. Phys. {\bf 208}
(1999) 413.

\bibitem{Witten-Thermal}  E. Witten, Adv. Theor. Math. Phys. {\bf 2} (1998)
505.

\bibitem{CHSW}  P. Candelas, G. T. Horowitz, A. Strominger and E. Witten,
Nucl.Phys. {\bf B258} (1985) 46.

\bibitem{GV}  M. Green and P. Vanhove, Phys. Lett. {\bf B408} (1997) 122.

\bibitem{Zwiebach}  B. Zwiebach, Phys.Lett. {\bf 156B} (1985) 315.

\bibitem{Zumino}  B. Zumino, Phys.Rep. {\bf 137} (1986) 109.

\bibitem{Lanczos}  C. Lanczos, Ann.Math. {\bf 39} (1938) 842.

\bibitem{Lovelock}  D. Lovelock, J.Math.Phys. {\bf 12} (1971) 498.

\bibitem{ACOTZ3+1}  R. Aros, M. Contreras, R. Olea, R. Troncoso and J.
Zanelli, Phys. Rev. Lett {\bf 84} (2000) 1647.

\bibitem{ACOTZ2n}  R. Aros, M. Contreras, R. Olea, R. Troncoso and J.
Zanelli, Phys. Rev. {\bf D62} (2000) 044002.

\bibitem{tz}  C. Teitelboim and J. Zanelli, Class. Quant. Grav. {\bf 4}
(1987) L125.

\bibitem{HTZ}  M. Henneaux, C. Teitelboim and J. Zanelli, {\em Gravity in
Higher Dimensions}, in SILARG V, M.Novello, (ed.), World Scientific,
Singapore, 1987{\bf ;} Phys. Rev. {\bf A36} (1987) 4417.

\bibitem{Wheeler}  J. T. Wheeler, Nucl. Phys. {\bf B268} (1986) 737; {\bf %
B273} (1986) 732.

\bibitem{Boulware-Deser}  D. G. Boulware and S. Deser, Phys. Rev. Lett. {\bf %
55} (1985) 2656.

\bibitem{HDG}  R. Troncoso and J. Zanelli, {\em Higher-dimensional
gravity, propagating torsion and AdS gauge invariance}. Class.
Quant. Grav. (2000) (to be published); Report CECS-PHY-99/12,
e-print: hep-th/9907109.

\bibitem{Achucarro-Townsend-Witten}  A. Ach\'{u}carro and P.K. Townsend,
Phys. Lett. {\bf B180} (1986) 89 ; E. Witten, Nucl. Phys. {\bf B 311} (1988)
46.

\bibitem{EGB}  In four dimensions, the integral of the Euler-Gauss-Bonnet
density is a topological invariant for compact manifolds without boundary.
In higher dimensions, this term gives rise to non- trivial contributions to
the field equations.

\bibitem{Spheritorsion}  In first order formalism, the field equations imply
the vanishing of torsion, except if $k=\left[ \frac{d-1}{2}\right] $ (BI and
CS), so that one is not necessarily forced to set $T^{a}=0$ from the start
\cite{HDG}. However, for static and spherically symmetric configurations,
equation (\ref{kTorsion}) implies that the torsion must vanish for these
cases as well.

\bibitem{BTZ}  M. Ba{\~{n}}ados, C. Teitelboim and J. Zanelli, Phys. Rev.
Lett. {\bf 69} (1992) 1849; M. Ba\~{n}ados, M. Henneaux, C.
Teitelboim, J. Zanelli, Phys. Rev.{\bf \ D} {\bf 48} (1993) 1506.

\bibitem{btz}  M. Ba\~{n}ados, C. Teitelboim and J. Zanelli, Phys. Rev. {\bf %
D49} (1994) 975.

\bibitem{MTZ}  C. Mart\'{\i }nez, C. Teitelboim and J. Zanelli, Phys. Rev.{\bf D61} (2000) 104013.

\bibitem{Regge-Teitelboim}  T. Regge and C. Teitelboim, Ann. Phys. (NY) {\bf %
88} (1974) 286.

\bibitem{Palais}  R. S. Palais, Comm. Math. Phys. {\bf 69} (1979) 19.

\bibitem{D-J-'tH}  S. Deser, R. Jackiw (MIT, LNS) and G. 't Hooft, Annals
Phys. {\bf 152} (1984) 220.

\bibitem{Muniain-Piriz}  J. P. Muniain and D. D. Piriz, Phys. Rev. {\bf D53}
(1996) 816.

\bibitem{Hawking-Page}  S. W. Hawking and D. Page, Comm. Math. Phys. {\bf 87}
(1983) 577.

\bibitem{Gibbons-Hawking}  G. W. Gibbons and S. W. Hawking, Phys. Rev. {\bf %
D15} (1977) 2752.

\bibitem{trz}  R. Troncoso and J. Zanelli, Phys. Rev. {\bf D58} (1998)
101703.

\bibitem{trz2}  R. Troncoso and J. Zanelli, Int. J. Theor. Phys.{\em \ }{\bf 38} (1999) 1193.

\bibitem{C-J-S}  E. Cremmer, B. Julia and J. Scherk, Phys. Lett. {\bf 76B}
(1978) 409.

\bibitem{B-D-H-S}  K. Bautier, S. Deser, M. Henneaux and D.Seminara, Phys.
Lett. {\bf B406} (1997) 49.

\bibitem{trz'}  R. Troncoso and J. Zanelli, {\em Chern-Simons Supergravities
with Off-Shell Local Superalgebras}{\it , }In 6th Meeting on
Quantum Mechanics of Fundamental Systems: Black Holes and
Structure of the Universe, Santiago, Chile, August 1997,
C.Teitelboim and J.Zanelli editors, World Scientific, Singapore
(1999). e-print: hep-th/9902003.

\bibitem{BTrZ}  M. Ba\~{n}ados, R. Troncoso and J. Zanelli, Phys. Rev. {\bf %
D54} (1996) 2605.

\bibitem{JJG}  M. Ba\~{n}ados, C. Teitelboim and J. Zanelli, {\em %
Lovelock-Born-Infeld Theory of Gravity}, in J.J.Giambiagi Festschrift,
edited by H. Falomir, R. E. Gamboa, P. Leal and F. Schaposnik (World
Scientific, Singapore, 1991).

\bibitem{chamslett}  A. H. Chamseddine, Phys. Lett. {\bf B233} (1989) 291%
{\bf ;} Nucl. Phys. {\bf B346} (1990) 213.
\end{references}
\end{document}